\documentclass[preprintnumbers,reprint,superscriptaddress,twocolumn]{revtex4}
\usepackage{textcomp}
\usepackage{graphicx}
\usepackage{amssymb}
\def\be{\begin{equation}}
\def\ee{\end{equation}}
\def\bea{\begin{eqnarray}}
\def\eea{\end{eqnarray}}
\def\met{\not{\!{\rm E}}_T}
\def\zp{Z^\prime}

\makeatletter

\begin{document}

\title{Observable $T_7$ Lepton Flavor Symmetry at the Large Hadron Collider}
\author{Qing-Hong Cao}
\affiliation{High Energy Division, Argonne National Laboratory, Argonne, 
Illinois
60439, USA}
\affiliation{Enrico Fermi Institute, University of Chicago, Chicago, Illinois
60637, USA}

\author{Shaaban Khalil}
\affiliation{Centre for Theoretical Physics, The British University in Egypt,
El Sherouk City, Postal No.~11837, P.O.~Box 43, Egypt}
\affiliation{Department of Mathematics, Ain Shams University,
Faculty of Science, Cairo 11566, Egypt}

\author{Ernest Ma}
\affiliation{Department of Physics and Astronomy, University of California,
Riverside, California 92521, USA}

\author{Hiroshi Okada}
\affiliation{Centre for Theoretical Physics, The British University in Egypt,
El Sherouk City, Postal No.~11837, P.O.~Box 43, Egypt}

\begin{abstract}
More often than not, models of flavor symmetry rely on the use of 
nonrenormalizable operators (in the guise of flavons) to accomplish 
the phenomenologically successful tribimaximal mixing of neutrinos. 
We show instead how a simple renormalizable two-parameter neutrino mass 
model of tribimaximal mixing can be constructed with the non-Abelian 
discrete symmetry $T_7$ and the gauging of $B-L$. This is also achieved 
without the addition of auxiliary symmetries and particles present in 
almost all other proposals. Most importantly, it is verifiable at the Large 
Hadron Collider.
\end{abstract}

\preprint{ANL-HEP-PR-10-50,~UCRHEP-T496} 

\maketitle

In 2001, the non-Abelian discrete symmetry $A_4$ was shown for the first 
time~\cite{mr01} to allow for the seemingly incompatible pattern that 
charged-lepton masses are all very different and yet a symmetry exists to 
predict the neutrino mixing matrix without knowing the individual 
neutrino masses. In 2004, it was shown for the first time~\cite{m04} that 
$A_4$ could also predict neutrino tribimaximal mixing with $\sin^2 2 
\theta_{atm} =1$ and $\tan^2 \theta_{sol} = 1/2$.  Since early 2005, when the 
solar angle in neutrino oscillations was revised by SNO~\cite{SNO} to 
$\tan^2 \theta_{sol} = 0.45 \pm 0.05$, this idea became widely accepted and 
the use of non-Abelian discrete symmetries~\cite{ikoost10} for understanding 
flavor has appeared in very many publications~\cite{af10}.  The two earliest 
papers~\cite{af05,bh05} after the SNO revision in 2005 both used $A_4$ and 
suggested two different two-parameter neutrino mass matrices, whereas 
the original proposal~\cite{m04} of 2004 had three parameters.

If the $3 \times 3$ Majorana neutrino mass matrix ${\cal M}_\nu$ is rotated 
by the Cabibbo-Wolfenstein unitary matrix~\cite{c78,w78}
\begin{equation}
U = {1 \over \sqrt{3}} \pmatrix{1 & 1 & 1 \cr 1 & \omega & \omega^2 \cr 1 
& \omega^2 & \omega},
\end{equation}
where $\omega = \exp(2 \pi i/3) = -1/2 + i\sqrt{3}/2$, then the tribimaximal 
form was shown~\cite{m04} to be
\begin{equation}
U {\cal M}_\nu U^T = \pmatrix{a+2b & 0 & 0 \cr 0 & a-b & d \cr 0 & d & a-b}.
\end{equation}
The two examples mentioned above are then~\cite{af05} $b=0$ and~\cite{bh05}
\begin{equation}
U {\cal M}_\nu U^T = \pmatrix{a-d^2/a & 0 & 0 \cr 0 & a & d \cr 0 & d & a}.
\end{equation}
However, these forms are only obtained at the expense of additional 
auxiliary symmetries and particles, and with the use of nonrenormalizable 
operators~\cite{af05}.

On the other hand, it has been shown recently~\cite{m10-1} that $A_4$ alone 
is sufficient to obtain $b=0$, if the alternative $A_4$ lepton assignments 
of Ref.~\cite{m06-1} are used instead of the original proposal of 
Ref.~\cite{mr01} and that neutrinos become massive through Higgs 
triplets~\cite{ms98} in a renormalizable model. Here we show how Eq.~(3) 
may be obtained by the canonical seesaw mechanism for neutrino 
mass, using the non-Abelian discrete symmetry $T_7$~\cite{lnr07} 
and gauging $B-L$~\cite{hkor08}, without the addition of auxiliary symmetries 
and particles or the use of nonrenormalizable operators.

Most importantly, our proposal is verifiable at the Large Hadron Collider 
(LHC). The predicted $\zp$ gauge boson will decay into scalars 
which support the $T_7$ symmetry.  Their subsequent decays 
into charged leptons will then reveal the predicted $T_7$ flavor structure 
used in obtaining neutrino tribimaximal mixing.

Since there are three families, non-Abelian discrete symmetries with 
irreducible three-dimensional representations are of special interest. 
The smallest group with a real \underline{3} representation is $A_4$ which 
has 12 elements.  The smallest group with a complex \underline{3} 
representation is $T_7$ which has 21 elements.  The group 
$\Delta(27)$~\cite{m06-2} is slightly bigger (27 elements) and 
also has a complex \underline{3} representation.  They are all subgroups 
of $SU(3)$.  The various irreducible representations of the three groups are
\begin{eqnarray}
A_4 &:& \underline{1}_i ~(i=1,2,3),~~\underline{3}; \\
T_7 &:& \underline{1}_i ~(i=1,2,3),~~\underline{3},~~\underline{\bar{3}}; \\ 
\Delta(27) &:& \underline{1}_i ~(i=1,2,3,4,5,6,7,8,9),~~\underline{3},
~~\underline{\bar{3}}. 
\end{eqnarray}
Their crucial differences are in the following group multiplications
\begin{eqnarray}
A_4 &:& \underline{3} \times \underline{3} = \sum_i \underline{1}_i + 
\underline{3} + \underline{3}; \\ 
T_7 &:& \underline{3} \times \underline{3} = \underline{3} + 
\underline{\bar{3}} + \underline{\bar{3}}, ~~~ 
\underline{\bar{3}} \times \underline{\bar{3}} = 
\underline{\bar{3}} + \underline{3} + \underline{3},\nonumber \\
& & \underline{3} \times \underline{\bar{3}} = 
\sum_i \underline{1}_i + \underline{3} + \underline{\bar{3}}; \\ 
\Delta(27) &:& \underline{3} \times \underline{3} = \underline{\bar{3}} + 
\underline{\bar{3}} + \underline{\bar{3}},  ~~~ \underline{\bar{3}} \times 
\underline{\bar{3}} = \underline{3} + \underline{3} + \underline{3}, 
\nonumber \\
& & \underline{3} \times \underline{\bar{3}} = \sum_i \underline{1}_i.
\end{eqnarray}
We will show that our $T_7$ model assignments cannot 
be replaced by either those of $A_4$ or $\Delta(27)$.

The finite group $T_7$ is generated by two noncommuting $3 \times 3$ 
matrices:
\begin{equation}
a = \pmatrix{\rho & 0 & 0 \cr 0 & \rho^2 & 0 \cr 0 & 0 & \rho^4}, ~~~ 
b = \pmatrix{0 & 1 & 0 \cr 0 & 0 & 1 \cr 1 & 0 & 0},
\end{equation}
where $\rho = \exp(2i\pi/7)$, so that $a^7=1$, $b^3=1$, and $ab=ba^4$. 
Let $\underline{3} = (x_1,x_2,x_3)$, and $\underline{\bar{3}} = (\bar{x}_1,
\bar{x}_2,\bar{x}_3)$, then their possible multiplications form the 
following $\underline{3}$ representations:
$(x_3y_3,x_1y_1,x_2y_2)$, $(x_2\bar{y}_1,x_3\bar{y}_2,x_1\bar{y}_3)$, 
$(\bar{x}_2\bar{y}_3 \pm \bar{x}_3\bar{y}_2,\bar{x}_3\bar{y}_1 \pm 
\bar{x}_1\bar{y}_3,\bar{x}_1\bar{y}_2 \pm \bar{x}_2\bar{y}_1)$,
and the following $\underline{\bar{3}}$ representations:
$(\bar{x}_3\bar{y}_3,\bar{x}_1\bar{y}_1,\bar{x}_2\bar{y}_2)$,  
$(x_1\bar{y}_2,x_2\bar{y}_3,x_3\bar{y}_1)$, 
$({x}_2{y}_3 \pm {x}_3{y}_2,{x}_3{y}_1 \pm {x}_1{y}_3,{x}_1{y}_2 \pm 
{x}_2{y}_1)$.
The combinations $x_1 \bar{y}_1 + \omega^{k-1} x_2 \bar{y}_2 + \omega^{2k-2} 
x_3 \bar{y}_3$ form the representations $\underline{1}_k$ for $k=1,2,3$ 
respectively.

Under $T_7$, let $L_i = (\nu,l)_i \sim \underline{3}$, $l^c_i \sim 
\underline{1}_i,~i=1,2,3$, $\Phi_i = (\phi^+,\phi^0)_i \sim \underline{3}$, 
which means that $\tilde{\Phi}_i = (\bar{\phi}^0,-\phi^-)_i \sim 
\underline{\bar{3}}$.  The Yukawa couplings $L_i l^c_j \tilde{\Phi}_k$ 
generate the charged-lepton mass matrix
\bea
m_l &=& \pmatrix{f_1 v_1 & f_2 v_1 & f_3 v_1 \cr f_1 v_2 & \omega^2 f_2 v_2 & 
\omega f_3 v_2 \cr f_1 v_3 & \omega f_2 v_3 & \omega^2 f_3 v_3} \nonumber\\
&=& 
{1 \over \sqrt{3}} \pmatrix{1 & 1 & 1 \cr 1 & \omega^2 & \omega \cr 1 & \omega 
& \omega^2} \pmatrix{f_1 & 0 & 0 \cr 0 & f_2 & 0 \cr 0 & 0 & f_3} ~v,
\eea
if $v_1 = v_2 = v_3 = v/\sqrt{3}$, as in the original $A_4$ 
proposal~\cite{mr01}.

Let $\nu^c_i \sim \underline{\bar{3}}$, then the Yukawa couplings 
$L_i \nu^c_j \Phi_k$ are allowed, with
\begin{equation}
m_D = f_D \pmatrix{0 & v_1 & 0 \cr 0 & 0 & v_2 \cr v_3 & 0 & 0} = 
{f_D v \over \sqrt{3}} \pmatrix{0 & 1 & 0 \cr 0 & 0 & 1 \cr 1 & 0 & 0},
\end{equation}
for $v_1 = v_2 = v_3 = v/\sqrt{3}$ which is already assumed for $m_l$.  
Note that 
$\Phi$ and $\tilde{\Phi}$ have $B-L=0$.  

Now add the neutral Higgs singlets $\chi_i \sim \underline{3}$ and $\eta_i 
\sim \underline{\bar{3}}$, both with $B-L=-2$.  Then there are two Yukawa 
invariants: $\nu^c_i \nu^c_j \chi_k$ and $\nu^c_i \nu^c_j \eta_k$ (which has 
to be symmetric in $i,j$).  Note that $\chi_i^* \sim \underline{\bar{3}}$ 
is not the same as $\eta_i \sim \underline{\bar{3}}$ because they have 
different $B-L$.  This means that both $B-L$ and the complexity of the 
$\underline{3}$ and $\underline{\bar{3}}$ representations in $T_7$ are 
required for this scenario. The heavy Majorana mass matrix for $\nu^c$ is then
\bea
M &=& h \pmatrix{u_2 & 0 & 0 \cr 0 & u_3 & 0 \cr 0 & 0 & u_1} + 
h' \pmatrix{0 & u'_3 & u'_2 \cr u'_3 & 0 & u'_1 \cr u'_2 & u'_1 & 0} 
\nonumber \\
&=& 
\pmatrix{A & 0 & B \cr 0 & A & 0 \cr B & 0 & A},
\eea
where $A = h u_1 = h u_2 = h u_3$ and $B = h' u'_2$, $u'_1 = u'_3 = 0$ have 
been assumed, i.e. $\chi_i$ breaks in the (1,1,1) direction, whereas $\eta_i$ 
breaks in the (0,1,0) direction.  This is the $Z_3-Z_2$ misalignment also 
used in $A_4$ models.

The seesaw neutrino mass matrix is now
\bea
&& m_\nu = - m_D M^{-1} m_D^T \nonumber \\
&& = {- f_D^2 v^2 \over 3A(A^2 - B^2)} \pmatrix{A^2-B^2 
& 0 & 0 \cr 0 & A^2 & -AB \cr 0 & -AB & A^2},
\eea
which has only two parameters and is identical to Eq.~(3). Detailed 
numerical analysis of this form was already done in Ref.~\cite{bh05}. 
Here we achieve the same result without the auxiliary $Z_4 \times Z_3$ 
symmetry and extra particles assumed there.  The key is that  
$\underline{\bar{3}} \times \underline{\bar{3}} \times \underline{3}$ 
is an invariant in $T_7$, but not in $\Delta(27)$, whereas $A_4$ cannot 
distinguish this from $\underline{\bar{3}} \times \underline{\bar{3}} 
\times \underline{\bar{3}}$ which yield two other invariants in $T_7$.

To realize the misalignment of $\langle \chi \rangle \sim (1,1,1)$ and 
$\langle \eta \rangle \sim (0,1,0)$, we need to choose the soft breaking 
terms in the Higgs potential consistent with these different residual 
symmetries~\cite{l07}.  However, the quartic terms $\chi_i^* \chi_j 
\eta_k^* \eta_l$ have several $T_7$ invariants, and most of them will 
destroy this pattern. To maintain the desired misalignment, this model has 
to be supersymmetrized. 

Consider $\chi \sim \underline{3}$ and $\eta \sim \underline{\bar{3}}$ as 
superfields with $B-L=-2$.  Add $\chi' \sim \underline{\bar{3}}$ and 
$\eta' \sim \underline{3}$ with $B-L=2$.  Then the superpotential 
contains the terms
\begin{equation}
W = f^\chi_{ijk} \nu^c_i \nu^c_j \chi_k + f^\eta_{ijk} \nu^c_i \nu^c_j \eta_k 
+ m_\chi \chi_i \chi'_i + m_\eta \eta_i \eta'_i,
\end{equation}
from which the $F$ terms of the Higgs potential are
\bea
V_F &=& |m_\chi \chi_i|^2 + |m_\eta \eta_i|^2 + |f^\chi_{ijk} \nu^c_i \nu^c_j 
+ m_\chi \chi'_k|^2 \nonumber \\
&+& |f^\eta_{ijk} \nu^c_i \nu^c_j + m_\eta \eta'_k|^2,
\eea
whereas the $D$ terms from $U(1)_{B-L}$ are
\begin{equation}
V_D = 2 g^2_{B-L} |\chi_i^* \chi_i + \eta_i^* \eta_i - {\chi'_i}^* \chi'_i 
- {\eta'_i}^* \eta'_i|^2.
\end{equation}
With the addition of bilinear soft terms $\chi_i^* \chi_i$, ${\chi'_i}^* 
\chi'_i$, $\chi_i \chi'_i + H.c.$, $\eta_2^* \eta_2$, ${\eta'_2}^* \eta'_2$, 
$\eta_2 \eta'_2 + H.c.$, $\eta_1^* \eta_1 + \eta_3^* \eta_3$, ${\eta'_1}^* 
\eta'_1 + {\eta'_3}^* \eta'_3$,  $\eta_1 \eta'_1 + \eta_3 \eta'_3 + H.c.$, and 
$(\chi_1 + \chi_2 + \chi_3) \eta'_2 + (\chi'_1 + \chi'_2 + \chi'_3) \eta_2 
+ H.c.$, which preserve $U(1)_{B-L}$ as they must, $T_7$ is broken 
with the desired pattern.

Flavor-changing leptonic interactions through Higgs exchange are present 
in this model, but they are suppressed by lepton masses, as in the original 
$A_4$ proposal~\cite{mr01}.  The set of three Higgs doublets $\Phi_i$ 
transforming as $\underline{3}$ under $T_7$ is rotated by $U$ of Eq.~(1) 
to form mass eigenstates $\phi_{0,1,2} \sim 1,\omega,\omega^2$ 
under the residual $Z_3$, where $\phi_0$ is identified as the one Higgs 
doublet (with $\langle \phi_0^0 \rangle = v$) of the Standard Model, 
with Yukawa couplings 
$v^{-1} [m_e \bar{e}_L e_R + m_\mu \bar{\mu}_L \mu_R + m_\tau 
\bar{\tau}_L \tau_R]$, which is of course flavor-conserving. 
The ``flavor-changing'' interactions of $\phi_{1,2}$ are then given by
\bea
{\cal L}_{int} = v^{-1} [m_\tau 
{\overline{L}_\mu}_L \tau_R + m_\mu {\overline{L}_e}_L \mu_R + 
m_e {\overline{L}_\tau}_L e_R] \phi_1 + && \nonumber \\ 
v^{-1} [m_\tau {\overline{L}_e}_L \tau_R 
+ m_\mu {\overline{L}_\tau}_L \mu_R + m_e {\overline{L}_\mu}_L 
e_R] \phi_2 + H.c. &&
\eea 
However, if the neutrino sector is ignored, a lepton flavor 
triality ($Z_3$ symmetry)~\cite{m09} exists here, under which
$e,\mu,\tau \sim 1,\omega^2,\omega$,  implying thus the decays $\tau^+ 
\to \mu^+ \mu^+ e^-$ and $\tau^+ \to e^+ e^+ \mu^-$, but no others.  
In particular, $\mu \to e \gamma$ is forbidden. Using
\[
B(\tau^+ \to \mu^+ \mu^+ e^-) = {m_\tau^2 m_\mu^2 (m_1^2 + m_2^2)^2 \over 
m_1^4 m_2^4} B(\tau \to \mu \nu \nu),
\]
the experimental upper limit of $2.3 \times 10^{-8}$ yields the 
bound~\cite{m09} $m_1 m_2/\sqrt{m_1^2+m_2^2} > 22$ GeV (174 GeV/$v$) 
on the masses of $\psi^0_{1,2} = (\phi^0_1 \pm \bar{\phi}^0_2)/\sqrt{2}$. 

Since the Higgs singlets $\chi$ and $\eta$ which support the neutrino 
tribimaximal mixing under $T_7$ also transform under $U(1)_{B-L}$, this 
model can be tested at the LHC by discovering the 
$Z^\prime_{B-L}(\equiv \zp)$ gauge boson. 
The partial decay rates of $\zp$ to the usual 
quarks and leptons are easily calculated.  Let $\Gamma_0 = g_{B-L}^2 m_{\zp}/12 
\pi$, then
$\Gamma_q = (6)(3)(1/3)^2\Gamma_0$, $\Gamma_l = (3)(-1)^2 \Gamma_0$,  
$\Gamma_\nu = (3)(-1)^2(1/2)\Gamma_0$.
As for $\zp \to {\psi}^0_{1,2} \bar{\psi}^0_{2,1}$, it has the effective 
partial rate $\Gamma_\psi \simeq (2)(-2)^2 \sin^4 \theta (1/4) \Gamma_0$, 
where $\sin \theta$ is an effective parameter accounting for the mixing of 
$\psi^0_{1,2}$ to $\chi$ and $\eta$ (with the help of a $B-L=0$ singlet $S_i 
\sim \underline{3}$). Using Eq.~(18), 
we find their signature decays to be given by
\begin{equation}
\psi^0_{1,2} \to \tau^+ \mu^-, ~ \tau^- e^+, ~~~  
\bar{\psi}^0_{1,2} \to \tau^- \mu^+, ~ \tau^+ e^-,
\end{equation}
resulting in $\zp$ leptonic final states such as $\tau^- \tau^- 
\mu^+ e^+$ for example.  In addition to being crucial for neutrino 
tribimaximal mixing to work under $T_7$, the $U(1)_{B-L}$ gauge symmetry 
is seen to provide also the means of verifying its predicted interactions. 
If the singlet neutrinos $\nu^c_i$ are light enough, they can also be 
produced by $\zp$ decay as discussed in Ref.~\cite{hkor08}.  The mass 
eigenstates of $\nu^c_i$ are given by Eq.~(13).  Their decays into 
$\phi_{1,2}$ and leptons, and the subsequent decays of $\phi_{1,2}$ 
to leptons (resulting in six leptons in the final state) will then give a 
complete picture of tribimaximal mixing in this model.

We now study in detail the process $q\bar{q} \to \zp \to \psi_1 \bar{\psi}_2 
+ \psi_2 \bar{\psi}_1$ (assuming $m_1=m_2$) 
with the subsequent decays $\psi \to \tau^- e^+$ and $\bar{\psi} \to \tau^- 
\mu^+$ at the LHC with $E_{cm} = 14$ TeV.  We consider only the leptonic 
decay modes of the $\tau^-$, with branching fraction $17.4\%$ to either 
$e^-$ or $\mu^-$.  The collider signature of such events is $e^+ \mu^+ 
\ell^- \ell^-$ plus missing energy, where $\ell=e,\mu$.  The dominant 
backgrounds yielding the same signature are the processes (generated by 
MadEvent/MadGraph~\cite{Maltoni:2002qb}):
\bea
&&WWZ: pp\to W^{+}W^{-}Z, W^{\pm}\to \ell^\pm \nu, Z\to \ell^{+}
\ell^{-},  \nonumber \\
&& ZZ: pp\to ZZ, Z\to \ell^+\ell^-,Z\to\tau^+\tau^-,
\tau^\pm\to \ell^{\pm} \nu\bar{\nu},\nonumber\\
&&t\bar{t}: pp\to t\bar{t}\to b(\to \ell^-)\bar{b}(\to \ell^+)W^{+}W^{-},\, 
W^{\pm}\to \ell^\pm\nu, \nonumber \\
&& Zb\bar{b}: pp\to Zb(\to \ell^-)\bar{b}(\to \ell^+),\, Z\to\ell^{+}\ell^{-}, 
\label{eq:smbkgd}
\eea
where $\ell=e,\mu$. 
Other SM backgrounds, e.g. $ZZZ$ and $WWWW$, occur at a negligible rate 
after kinematic cuts, and are not shown here.  
We require no jet tagging and consider only events with both $e^+$ and 
$\mu^+$ in the final state. 
The first two processes are the irreducible background, while the last two 
are reducible as they only contribute when some tagged particles escape 
detection, carrying away small transverse momentum ($p_{T}$) or 
falling out of the detector rapidity coverage.

\begin{figure}
\begin{center}
\includegraphics[scale=0.37]{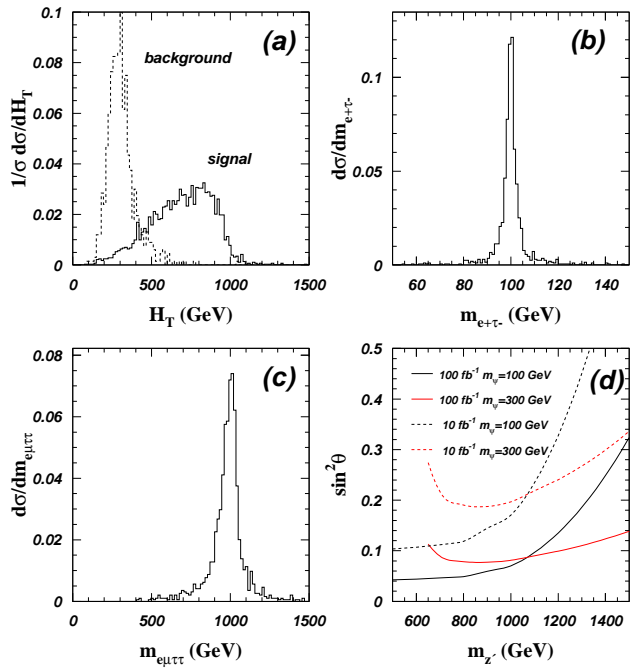}
\caption{ (a) Normalized distribution of $H_T$; (b) Distribution of the 
invariant mass of the $e^+$ and reconstructed $\tau^-$pair; (c) Distribution 
of the mass of the reconstructed $\zp$; (d) The $5\sigma$ significance contours
in the plane of $m_{\zp}$ and $\sin^2\theta$. 
\label{fig:lhc} }
\end{center}
\end{figure}

Our benchmark points are chosen as follows:
$m_{\zp} = 1000~(1500)~{\rm GeV},~m_{\psi} = 100~(300)~{\rm GeV}$, 
$g_{B-L} = g = e/\sin \theta_W$, and $\sin^2\theta = 0.2$.
In our analysis all events are required to pass the following
{\it basic} acceptance cuts:
\bea
& & p_{T,\,\ell}^{(1,2)}\geq 50\,{\rm GeV},
~p_{T,\,\ell}^{(3,4)}\geq 20\,{\rm GeV},
~\left|\eta_{\ell}\right|\leq2.5,
\nonumber \\
& & \Delta R_{\ell\ell^\prime}\geq 0.4,~\met > 30~\rm{GeV},
\label{eq:cut}
\eea
where (1-4) in the superscript index is the $p_T$ order of the charged leptons.
$\Delta R_{ij}$ is the separation in the azimuthal angle ($\phi$) 
- pseudorapidity ($\eta$) plane between $i$ and $j$, defined as 
$\Delta R_{ij}\equiv \sqrt{\left(\eta_{i}-\eta_{j}\right)^{2}+\left(\phi_{i}-
\phi_{j}\right)^{2}}$.  We also model detector resolution effects by smearing 
the final-state energy. To further suppress the SM backgrounds, we demand
\begin{equation}
H_T \equiv \sum_i p_{T,~i} + \met > 300~{\rm GeV}, \label{eq:ht}
\end{equation}
where $i$ denotes the visible particles. 
Figure~\ref{fig:lhc}(a) shows the normalized $H_T$ distribution of both signal
and background before the $H_T$ cut. The signal spectrum exhibits an endpoint 
around the mass of $\zp$, about 1~TeV, but with a long tail due to the 
$\zp$ width and detector smearing effects. Table~\ref{tab:cut} displays
the signal and background cross sections (fb) before and after cuts. 
The cut acceptance ($\mathcal{A}_{cut}$) increases with $m_\psi$ 
as heavy scalar decay generates hard leptons and large $H_T$. 
For a light $\psi$ and a heavy $\zp$ (e.g. the benchmark B), $\mathcal{A}_{cut}$ 
decreases as the two charged leptons from the light scalar decay are 
very much parallel and 
fail the $\Delta R$ separation cuts.

\begin{table}
\caption{Signal and background cross sections (fb) before and after cuts 
for four $(m_{\zp},m_\psi)$ (GeV) benchmark points: (A) (1000,100), (B) (1500,100),
(C) (1000,300), and (D) (1500,300).  
The ``no cut'' rates correspond to all leptonic decay modes of $\tau^-$
after $e^+\mu^+$ identification, 
the ``basic cut'' and ``$H_T$ cut'' rates are obtained after imposing 
Eq.~(\ref{eq:cut}) and Eq.~(\ref{eq:ht}), respectively, 
whereas the ``$x_{\tau_i}>0$'' rates are obtained after the $\tau^-$ 
reconstruction cuts. The bottom row shows the cut acceptance ($\mathcal{A}_{cut}$).
\label{tab:cut}}
\begin{center}
\begin{tabular}{c|cccc|cccc}
\hline 
              & (A)  & (B)    & (C) & (D) & $t\bar{t}$& $WWZ$ & $ZZ$ & $Zb\bar{b}$ \tabularnewline
\hline
no cut        &5.14  & 0.98   &2.57 & 0.72 & 1.22   & 0.21   & 27.11   & 2.99     \tabularnewline
basic cut     &1.46  & 0.066  &1.05 & 0.36 & 0.16   & 0.02   & 0.0052  & 0.024    \tabularnewline
$H_T$ cut     &1.41  & 0.065  &1.04 & 0.36 & 0.08   & 0.006  & 0.0     & 0.0      \tabularnewline
$x_{\tau}>0$  &0.69  & 0.032  &0.52 & 0.18 & 0.015  & 0.002  & 0.0     & 0.0      \tabularnewline
\hline
$\mathcal{A}_{cut}$&13.4\%  & 3.2\%  & 20\%& 25\% & 1.2\%  & 1\%  & 0.0     & 0.0      \tabularnewline
\hline
\end{tabular}
\end{center}
\end{table}

To reconstruct the scalar $\psi$, we adopt the collinear approximation 
that the charged lepton and neutrinos from $\tau$ decays are parallel 
due to the large boost of the $\tau$. Such a condition is satisfied to an 
excellent degree because the $\tau$ leptons originate from a heavy scalar 
decay in the signal event.  Denoting by $x_{\tau_i}$ the fraction of the parent 
$\tau$ energy which each observable decay particle carries, the transverse 
momentum vectors are related by~\cite{Rainwater:1998kj}
\be
\vec{\not{\!E}}_T = \left(1/x_{\tau_1} - 1 \right)\vec{p}_1
+\left(1/x_{\tau_2}-1\right)\vec{p}_2. \label{eq:taurec}
\ee
When the decay products are not back-to-back, Eq.~(\ref{eq:taurec}) gives two 
conditions for $x_{\tau_i}$ with the $\tau$ momenta as $\vec{p}_1/x_{\tau_1}$ 
and $\vec{p}_{2}/x_{\tau_2}$, respectively. We further require the calculated 
$x_{\tau_i}$ to be positive to remove the unphysical solutions.
There are two possible combinations of $e^+\ell^-$ clusters 
for reconstructing the scalar $\psi$ and gauge boson $\zp$. To 
choose the correct combination, we require the $e^+\ell^-$ pairing to be 
such that $\Delta R_{e^+\ell^-}$ is minimized.
The mass spectra of the reconstructed $\psi$ 
and $\zp$ are plotted in 
Fig.~\ref{fig:lhc}(b) and (c), respectively, which clearly 
display sharp peaks around $m_\psi$ and $m_{Z^\prime}$.  
In Fig.~\ref{fig:lhc}(d), we show the $5\sigma$ discovery contours 
in the plane of $m_{\zp}$ and $\sin^2\theta$ by requiring 8.5 (5) signal 
events for an integration luminosity of $100~(10)~{\rm fb}^{-1}$ 
respectively. The regions above those curves 
are good for discovery.

In the quark sector, if we use $Q_i = (u,d)_i \sim \underline{3}$ and 
$u^c_i,d^c_i \sim \underline{1}_i, i =1,2,3$ as we assume for the charged 
leptons, we again obtain arbitrary quark masses, but no mixing. 
To have realistic mixing angles, the residual $Z_3$ symmetry has to be broken.  

Non-Abelian discrete symmetries have been successful in explaining the 
tribimaximal mixing of neutrinos, but not their masses.  However, they are 
very difficult to 
test experimentally. In this paper, by combining $T_7$ and $U(1)_{B-L}$, 
we show how a simple renormalizable two-parameter neutrino mass model of 
tribimaximal mixing can be constructed, with verifiable experimental 
predictions.  The key is the possible discovery of $\zp$ at the TeV 
scale, which then decays into neutral Higgs scalars, whose subsequent 
exclusive decays into charged leptons have a distinct flavor pattern 
which may be observable at the LHC.

The work of Q.H.C. is supported in part by the U.~S.~Dept.~of Energy  
Grant No.~DE-AC02-06CH11357 and in part 
by the Argonne National Lab.~and Univ.~of Chicago Joint Theory Institute 
Grant No.~03921-07-137. 
The work of S.K. and H.O. is supported in part by the Science and Technology 
Development Fund (STDF) Project ID 437 and the ICTP Project ID 30.
The work of E.M. is supported in part by the U.~S.~Dept.~of Energy  
Grant No. DE-FG03-94ER40837.

\bibliographystyle{unsrt}

\end{document}